%% file: PG21-egocentric_VR_sub.tex
\documentclass{egpubl}
\usepackage{pg2021}

\SpecialIssuePaper         

\usepackage[T1]{fontenc}
\usepackage{dfadobe}  

\usepackage{cite}  

\usepackage{microtype}                 
\PassOptionsToPackage{warn}{textcomp}  
\usepackage{textcomp}                  
\usepackage{mathptmx}                  
\usepackage{times}                     
\usepackage{tabu}                      
\usepackage{booktabs}                  %

\newcommand{\johannes}[1]{{\color{black}{#1}}}
\newcommand{\alessio}[1]{{\color{black}{#1}}}
\newcommand{\rev}[1]{{\color{black}{#1}}}

\usepackage{subfig}
\usepackage{tabularx}
\usepackage{xspace}

\BibtexOrBiblatex
\electronicVersion
\PrintedOrElectronic
\ifpdf \usepackage[pdftex]{graphicx} \pdfcompresslevel=9
\else \usepackage[dvips]{graphicx} \fi

\usepackage{egweblnk} 
\usepackage{xcolor}
\usepackage{pdfpages}

\newcommand{\egotwo}{Ego-Highlight\xspace}
\newcommand{\egothree}{Ego-Bubble\xspace}

\title[Egocentric Network Exploration for IA]%
      {Egocentric Network Exploration for Immersive Analytics}
      
\author[J. Sorger et al.]
{\parbox{\textwidth}{\centering J. Sorger$^{1}$\orcid{0000-0002-5212-8941},
        A. Arleo$^{2}$\orcid{0000-0003-2008-3651},
        P. K\'{a}n$^{2}$\orcid{0000-0001-7437-9955},
        W. Knecht$^{1}$\orcid{0000-0001-9530-7598},
        and M. Waldner$^{2}$\orcid{0000-0003-1387-5132}
        }
        \\
{\parbox{\textwidth}{\centering $^1$Complexity Science Hub, Vienna, Austria\\
        $^2$TU Wien, Vienna, Austria
       }
}
}

\begin{document}

\null

\includepdf[fitpaper=true, pages=-]{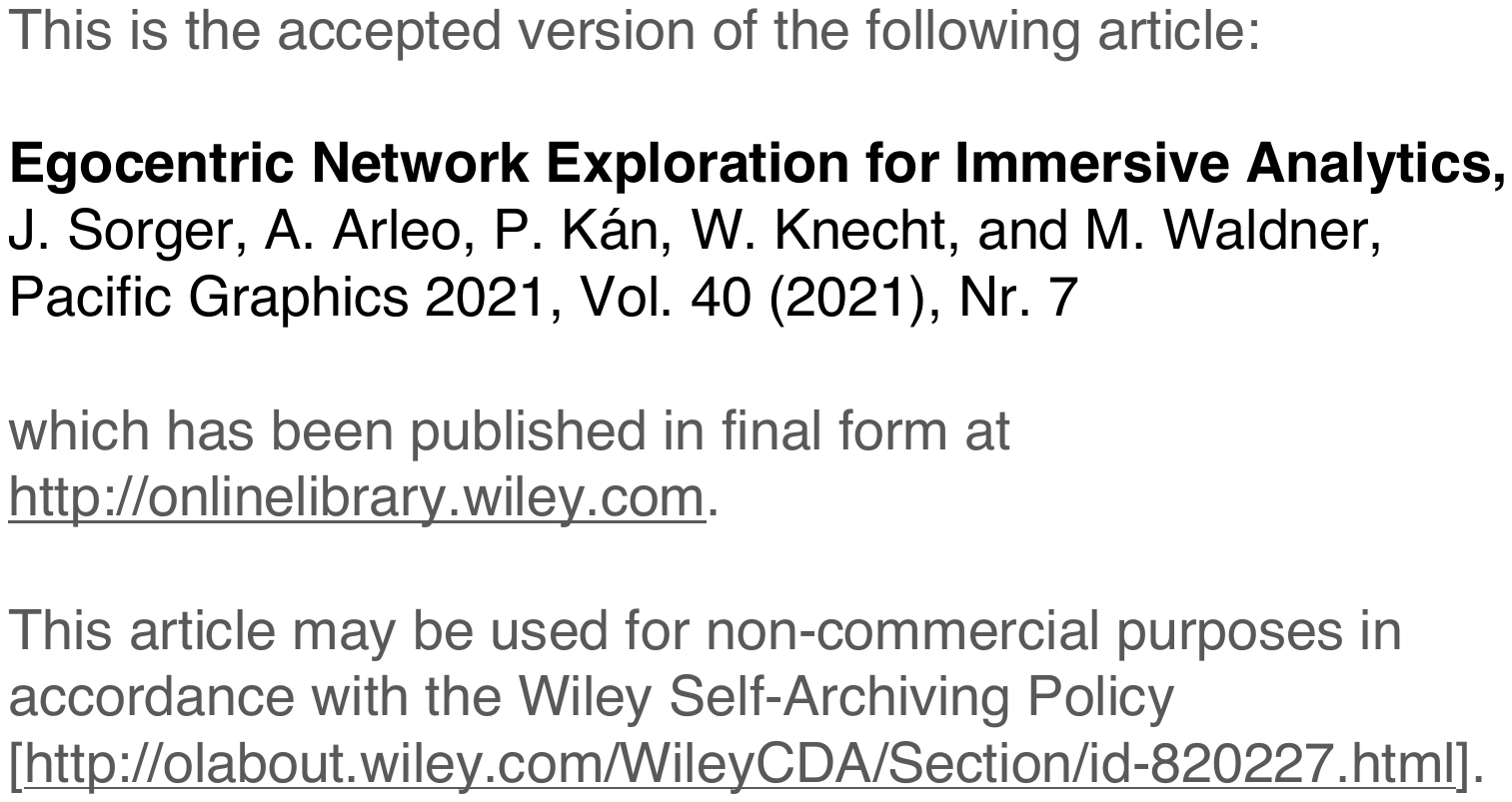}

\teaser{
  \centering
    \includegraphics[width=\linewidth]{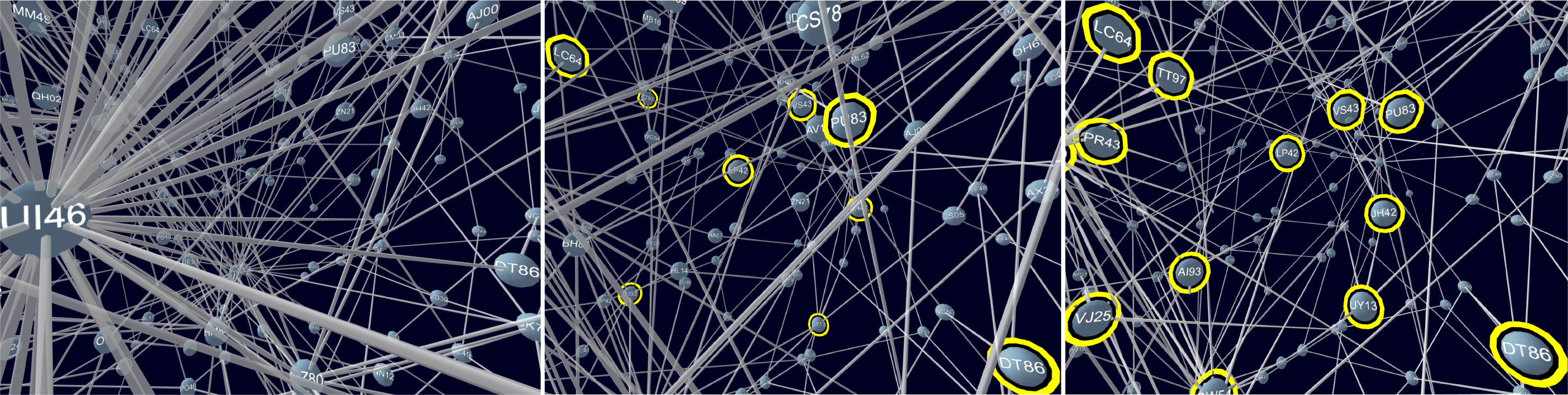}
    \caption{Exploring the local neighborhood of a hub node (`UI46') in three immersive interfaces. Baseline (left): the user inspects `UI46' from close-distance. \egotwo (middle): the user takes the position of `UI46'; direct neighbors are highlighted in yellow. \egothree (right): direct neighbors are arranged equidistantly to the user; edges that cross the user's bounding sphere are clipped.}
   \label{fig:teaser}
}

\maketitle
\begin{abstract}
    To exploit the potential of immersive network analytics for engaging and effective exploration, we \rev{promote} the metaphor of ``egocentrism'', where data depiction and interaction are adapted to the perspective of the user within a 3D network. Egocentrism has the potential to overcome some of the inherent downsides of virtual environments, e.g., visual clutter and cyber-sickness.
    To investigate the effect of this metaphor on immersive network exploration, we designed and evaluated interfaces of varying degrees of egocentrism. 
    In a user study, we evaluated the effect of these interfaces on visual search tasks, efficiency of network traversal, spatial orientation, as well as cyber-sickness. Results show that a simple egocentric interface considerably improves visual search efficiency and navigation performance, yet does not decrease spatial orientation or increase cyber-sickness. An occlusion-free \egothree view of the neighborhood only marginally improves the user's performance. We tie our findings together in an open online tool for egocentric network exploration, providing actionable insights on the benefits of \rev{the egocentric network exploration} metaphor.


\end{abstract}
\section{Introduction}
Immersive analytics (IA) investigates \emph{``engaging and embodied analysis tools to support data understanding and decision making''}~\cite{marriott2018immersive}. 
Particularly in network analytics, prior research has empirically demonstrated many benefits of three-dimensional immersive displays over traditional two-dimensional graph visualization methods~\cite{ware1996evaluating,belcher2003using,ware2008visualizing, halpin2008exploring,greffard2011visual,kwon2016study, kotlarek2020study}.
In virtual reality (VR), users can interact with the data as if they were a physical entity that they can approach and inspect. They can navigate \textit{within} a network instead of just looking \textit{at} it. Immersion in VR can reduce visual clutter~\cite{bowman2007virtual} and provide insightful complementary perspectives on a network~\cite{sorger2019immersive}.
Room-scale immersive networks can lead to more user engagement compared to a table-scale network coupled with a zooming interface~\cite{yang2020embodied}.
Indeed, there are numerous examples of immersive network analytics interfaces presented in literature that allow users to ``dive into'' and subsequently walk or fly through a 3D network~\cite{zielasko2016evaluation,drogemuller2017vrige,erra2019virtual,sorger2019immersive,yang2020embodied}.
While being surrounded by a room-sized network can be interesting, it can also be problematic. 
Navigation and interaction in abstract 3D data spaces is more challenging compared to 2D~\cite{marriott2018immersive_chapter}, because abstract data spaces do not have a natural scale and orientation~\cite{yang2020embodied}. An immersive perspective can also require more positional changes by the user to resolve occlusions and blind spots~\cite{kraus2019impact} 
-- a challenge for data analysts who prefer to seamlessly enter and leave the virtual environment while remaining at their desks~\cite{zielasko2017remain}. Flying through a cluttered visualization can then lead to cyber-sickness~\cite{sorger2019immersive} caused by the sensory conflicts between the static user and the visual sensation of a moving environment~\cite{laviola2000discussion}. 
We believe that with careful design around the user, which we refer to as \textit{egocentric}, these challenges can be alleviated while still exploiting the beneficial capabilities of VR.

In graph visualization, egocentric network analysis techniques support users in analyzing the part of the social network that is closely related to  the node that represents themselves in the network~\cite{fisher2005using,GOLBECK20139}. We translate this principle to immersive network analytics: by applying the egocentric network metaphor to 3D node-link diagrams, users can take the perspective of any node -- they can ``be the node'' -- and thereby gain an optimized view of their local neighborhood, i.e., data points that are closely related to ``them''. Associating the user with a dedicated node allows us to optimize the visibility of the local neighborhood, thereby resolving occlusion problems and removing visual clutter. The egocentric approach also has the potential for more efficient navigation within the network: by using data points (nodes) as anchors for exploration, users can change their egocentric perspective on the network by ``jumping'' from node to node. Yet, the restriction of the user's movement and the introduction of local adaptions of the data representation may cause disorientation during network traversal. 
In this paper, we set out to systematically explore the trade-offs that the concept of egocentricsm introduces to immersive network exploration.
Understanding the relevance of these trade-offs in analysis tasks is crucial to conceive methods that provide an effective and pleasant exploration experience to users. 
The contributions of this paper are:
\begin{itemize}
    \item \johannes{\rev{Egocentric detail views} as a new metaphor for immersive network analytics, exemplified by two novel VR interfaces with varying degrees of egocentrism; }
    \item Results of a controlled user study systematically investigating the effect of the proposed interfaces on visual search efficiency, navigation performance, spatial orientation, cyber-sickness, and user preference;
    \item A publicly available web application for immersive egocentric network exploration synthesizing the insights from the study.
\end{itemize}

\input{relatedwork}

\section{Egocentrism in Immersive Network Analytics}

We promote \textit{egocentrism} in the context of immersive analytics to better exploit the capabilities of VR technology for locally optimized visual representation and interaction capabilities. 
In contrast to traditional 2D viewing and interaction modalities, a VR application can be made aware of the users' position and scale in the virtual scene, their field of view, and potentially even focus of attention with respect to the data they are exploring.
Adaptions to an interface to support egocentrism can thus affect  navigation, interaction, rendering, or the spatial layout of the data.

\begin{figure}[b]
    \centering
    \includegraphics[width=0.85\linewidth]{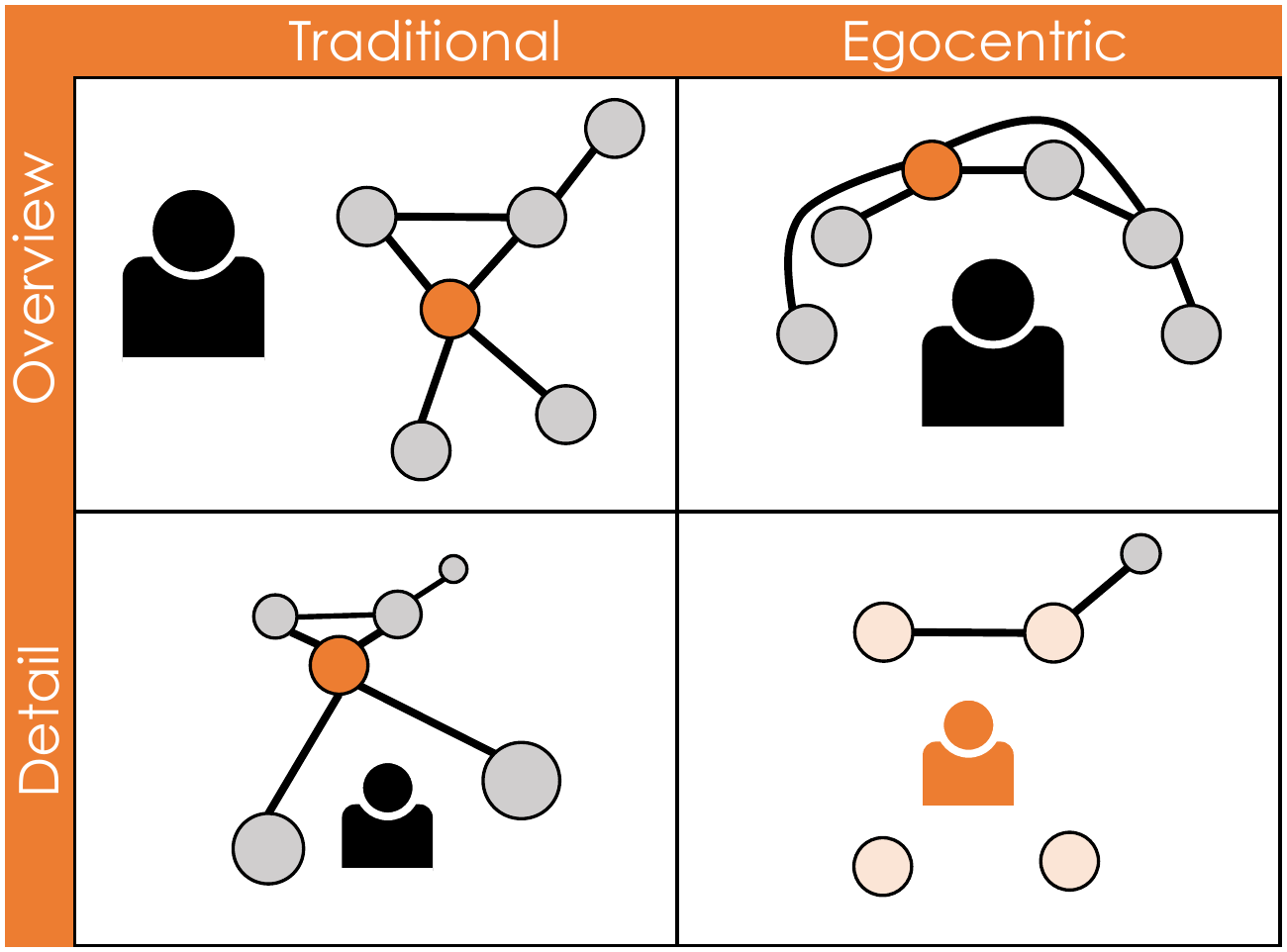}\vspace{-2mm}
    \caption{Categorization of network exploration techniques. The node used for the egocentric detail perspective (bottom right) is colored orange in the other networks. }
    \label{fig:categorization}
\end{figure}

\begin{figure}[t]
    \centering
    \includegraphics[width=0.95\linewidth]{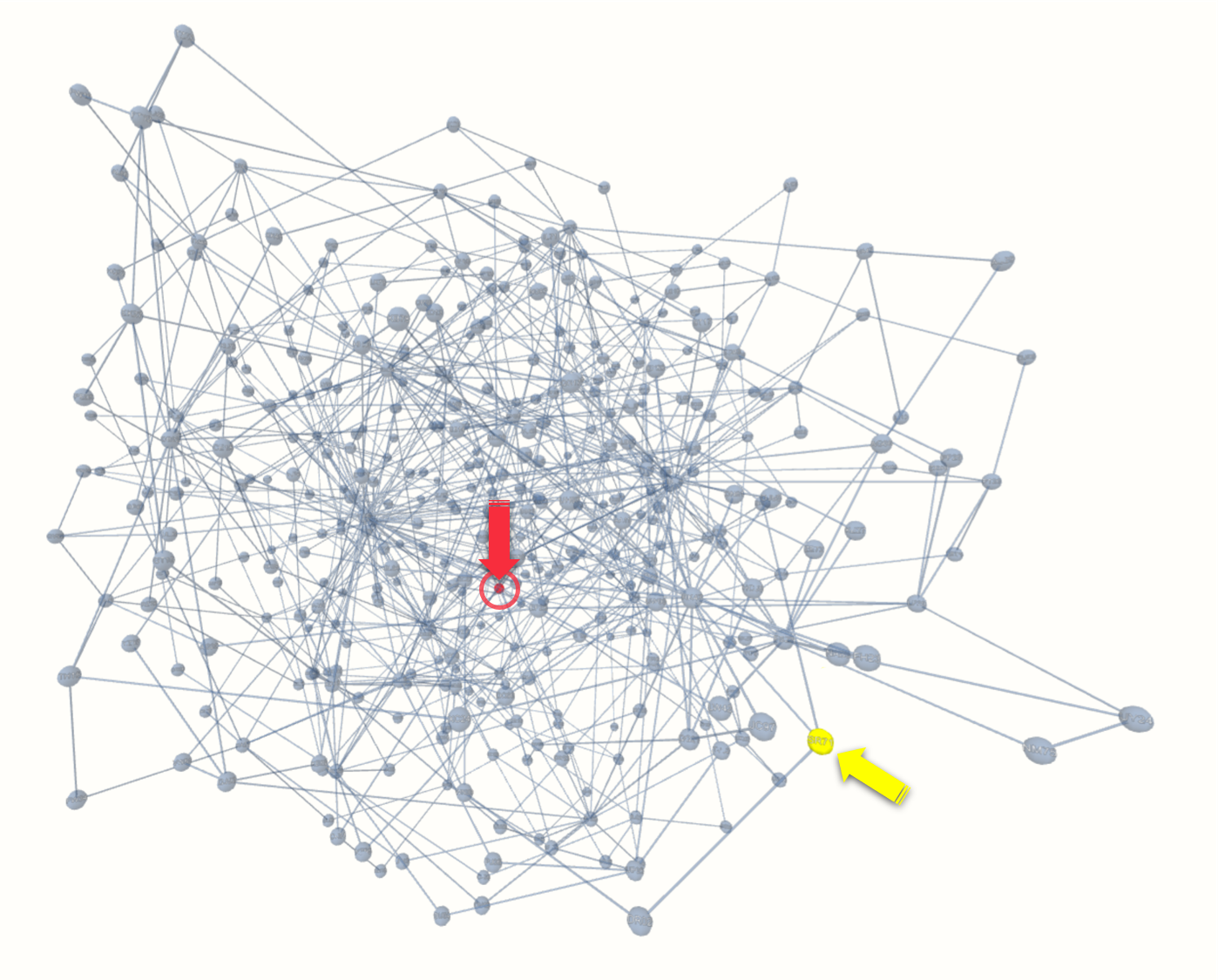}\vspace{-2mm}
    \caption{Overview perspective of a large graph used in the study: the starting condition of the Find Path (FiP) task, showing the start (yellow) and end (red) node of the path. Hubs and outliers can be spotted more easily here, but local connectivity is hard to explore.}
    \label{fig:overview}
\end{figure}

We categorize network exploration techniques in VR along two orthogonal axes, as illustrated in Figure~\ref{fig:categorization}: On the vertical axis, we differentiate whether users explore the network from an external/global ``overview'' perspective, from where the entire graph should be visible (see Fig.~\ref{fig:overview}), or from an internal/local ``detail'' perspective within the network (see Fig.~\ref{fig:teaser}).
On the horizontal axis, we distinguish whether or not the network visualization and the interaction techniques follow an egocentric approach, i.e., are optimized to the user's position and orientation in the virtual environment.
The spherical layout proposed by Kwon et al.~\cite{kwon2015spherical}, for instance, can be classified as an egocentric overview interface\johannes{, as it offers an overview that takes the user's position into account.} The overview+detail interface by Sorger et al.~\cite{sorger2019immersive} and the table- and room-scale visualizations by Yang et al.~\cite{yang2020embodied} combine traditional overview and detail views \rev{(note that Yang et al.~\cite{yang2020embodied} refer to the traditional overview (Fig. \ref{fig:categorization} top left) as ``exocentric'' and to the traditional detail view (bottom left) as ``egocentric view'').
To the best of our knowledge, \emph{egocentric detail views} (Fig. \ref{fig:categorization} bottom right) have not been investigated yet. They are the focus of this paper.}
We argue that an overview alone will lead to little benefit of using VR. Especially when viewing large networks from an external perspective, nodes and clusters in the center could be occluded. Having a detail view alone, on the other side, lacks the necessary overview to identify interesting regions in the network topology or to put the local environment into a global context. In line with the information seeking mantra, egocentric network exploration interfaces should therefore be coupled with an overview perspective that can serve as a stable reference for analysts switching between immersive and traditional desktop analytics~\cite{sorger2019immersive}. 

\section{A Study on Egocentrism for Immersive Network Exploration}
\label{sec:ego}
We expect that network exploration interfaces that support egocentrism can lead to a trade-off affecting \rev{several} factors, such as visual search efficiency, navigation performance, spatial orientation, and cyber-sickness.
To quantify these trade-offs, as well as to assess the overall benefits and limitations of egocentrism for network exploration in IA, we conducted a controlled user study where we compared the effectiveness and efficiency of typical low-level network analytics tasks (see Sec.~\ref{sec:tasks}) between a traditional interface (the "baseline") and two interfaces with an increasing level of egocentrism (``\egotwo'' and ``\egothree''). 



Our proposed interfaces target traditional data analysts that are accustomed to network analysis in 2D and thus are also aware of the limitations of conventional two-dimensional approaches.
Considering the space requirements in conventional offices, our interfaces are designed for desktop-bound VR setups rather than room-scale ones. Free room-scale movement is possible but not required for navigating the networks in VR. This also facilitates frequent switching between desktop work and immersive analytics, which is required for data analysis~\cite{zielasko2017remain}. 
The types of networks commonly analyzed in network analytics often correspond to 
small-world networks, e.g., social or financial networks, co-morbidity networks, or metabolic pathways (see Sec.~\ref{sec:data}). 

\subsection{Egocentric Network Exploration Interfaces}
\label{sec:general}
The three interfaces that we investigate in our study (the baseline interface and the two egocentric interfaces), represent a continuum of egocentrism, i.e., from no alterations in the baseline, over adaptions to the representation (\egotwo), to local layout adaptions (\egothree). 
This continuum also leads to tension between two conflicting aspects: 1) freedom of navigation vs.~ease of navigation and
2) layout fidelity vs.~legibility.
Both aspects are assumed to have an impact on visual search efficiency, navigation performance, spatial orientation, and cyber-sickness for topological graph analysis tasks, as discussed in Section~\ref{sec:tasks}. 



The \textbf{baseline} interface displays a traditional 3D representation of the force-directed node-link diagram and features 6 \johannes{degrees-of-freedom (DoF)} free-fly navigation (see Fig.~\ref{fig:teaser}, left).
Nodes are displayed as spheres, and links between the nodes are represented as tubes.
The interface serves as a baseline condition in our experiments, so no egocentric adaptions are implemented.
Navigation is handled via touch-sensitive input on the two axes of a trackpad, like the one on the HTC Vive controller.
The user flies along and perpendicular to his/her current viewing direction, as defined by the HMD view-vector -- just like moving and strafing in first-person ego-shooters.
Since the layout remains unaltered, nodes and edges can serve as stable landmarks to support spatial orientation.
However, edges can easily obstruct the user's view, especially in close proximity to nodes.


\textbf{\egotwo}: 
In line with the principle of egocentric network analytics \cite{fisher2005using}, the user can associate him/herself with a node (i.e., the user-node), and view the network from this particular perspective. 
The user-node, as an explicit user defined reference point, allows us to introduce egocentric adaptions to the interface: we explicitly highlight direct neighbor nodes with a yellow halo (see Fig.~\ref{fig:teaser}, middle). This in turn, allows us to remove the now redundant links to direct neighbors
to reduce visual clutter.
This metaphor can further simplify navigation by restricting movement to jumps between nodes.
By selecting a node with the controller's laser pointer and initiating a jump with the trigger button,
the user is automatically transported to the center of the selected node with an animated translation of the user's position. The user maintains control of the camera orientation throughout, as it is linked to the HMD.
This adaption trades flexibility in navigation for ease of use, as navigation no longer requires constant user input. 
Due to the nature of the force-directed layout, some neighbor nodes can be far away from the user-node, and might still be occluded through nodes and links that are not part of the direct neighborhood. We aim to further address this issue with additional egocentric adaptions in the next interface.

The \textbf{\egothree} interface extends the \egotwo interface by locally optimizing the network layout in respect to the current user-node. In order to further alleviate potential occlusion in the user's \johannes{local view, direct neighbor nodes are shifted to be evenly distributed around the user. We use the Fibonacci sphere algorithm on a bounding sphere centered around the user's head.}
Edges that cross the sphere are clipped (see Fig.~\ref{fig:teaser}, right) to avoid view obstructions through links passing by the vicinity of the user.
The overall layout is thereby still preserved and only locally distorted, aiming to facilitate the local neighborhood analysis of the current user-node. 
This type of node displacement is inspired by well-established fish-eye techniques~\cite{sarkar1994graphical}, employed in 2D visualizations of large graphs~\cite{abello2004visualizing}.
A trade-off for this cleaner display of local neighborhoods is a potential negative impact on the user's orientation through the displacement of node positions.
We therefore animate the local node displacement when jumping between nodes,  resulting in a morphing effect during navigation that is designed to help users keep and update their mental map, partially remediating the potential loss of orientation. 



\subsection{Hypotheses}
\label{sec:hypo}

We tested six hypotheses concerning the potential benefits and negative side-effects of egocentrism in immersive network exploration: 

\textbf{H1}: \emph{\textbf{Local visual search} is the most efficient in the \egothree condition and the least efficient in the baseline condition.}
Visual search is the task to spot a known target within a set of distractor items~\cite{wolfe1989guided}. 
If targets in the user's local neighborhood (i.e., neighboring nodes or edges) are occluded or outside the user's field of view, visual search will be negatively affected. 
We therefore expect to observe a considerable increase of visual search efficiency during local neighborhood exploration using egocentric interfaces that optimize the visibility of the user's neighborhood. 

\textbf{H2}: \emph{\textbf{Global visual search} is the most efficient using the baseline.}
If the user's task is to search for more global structures, such as a path along multiple nodes through the network, it might be required to freely adjust the viewpoint to improve visibility. In the baseline the graph representation remains stable and navigation is not restricted to node positions. Visual search for global structures therefore can be expected to be more efficient using the baseline interface. 

\textbf{H3}: \emph{\textbf{Navigation performance} is higher with egocentric network traversal than with flying in the baseline.}
Navigation performance is an important factor for the overall task efficiency when traversing a network in VR, e.g., when following a path. Traditional flying like in the baseline can lead to cyber-sickness~\cite{langbehn2018evaluation}, with symptoms worsening as the movement pace increases~\cite{so2001effects}. In an informal pilot study, we varied the animation speed between consecutive jumps and flying, and asked users to report their subjective well-being. \johannes{As a result, the maximal flying speed is kept slow to maintain acceptable cyber-sickness rates. Contrarily, the jumping speed always measures three seconds (including ease-in/out de/accelaration).} \rev{Using the baseline, flying along a path of five nodes in one of our networks (see Sec.~\ref{sec:data}) requires around 25 seconds, while the jump animations along five consecutive path nodes using the egocentric conditions results in around 15 seconds overall travel time.}
It can therefore be expected that navigation performance increases \rev{by a factor of $\sim1.7$} for the egocentric network interfaces. This is also \rev{suggested} by prior work, which shows that jumping leads to significantly faster travel times~\cite{weissker2018spatial}. However, besides the motoric aspect, navigation consists also of wayfinding, which is the cognitive and stationary aspect of navigation~\cite{darken2014spatial}. We believe, however, that a potentially negative impact of the egocentric interface on wayfinding will not affect the clearly superior motoric aspect.

\textbf{H4}: \emph{\textbf{Orientation} is affected negatively through egocentric adaptions, i.e., suffers least in the baseline, and the most using \egothree.}
A study by Drogemuller et al.~\cite{drogemuller2018evaluating} has shown that flying causes less loss of orientation than teleportation when traversing a 3D network in VR. Orientation loss can also be observed for jumping, but to a lesser extent~\cite{weissker2018spatial}. In contrast to these studies, the appearance of our scene changes after a jump, which can make it harder to keep oriented. Since the \egothree interface 
also changes the layout of the nodes between jumps, we assume that users experience most orientation loss in this condition.

\textbf{H5}: \emph{\textbf{Cybersickness} symptoms are less severe when using \egotwo compared to the baseline and \egothree}. 
Potential causes for cyber-sickness are, among others, sensory conflicts of a stationary user flying through a virtual environment~\cite{laviola2000discussion}. 
There is prior evidence that jumping is less prone to cyber-sickness symptoms than omni-directional flying \cite{langbehn2018evaluation}. However, this is based on the assumption that the scene is static during the transition. In the \egothree condition, the position of the surrounding nodes also changes during the animated jump, which could have a negative impact on the subjective well-being of participants compared to \egotwo. 

\textbf{H6}: \emph{\textbf{User preference} will be significantly higher for the egocentric interfaces than for the baseline}. 
We assume that the expected benefits with respect to visual search and navigation performance will outweigh the expected draw-backs so that users prefer egocentric network exploration interfaces over the baseline. We also assume that their perceived work load will be lower.

\subsection{Tasks}
\label{sec:tasks}

To enable accurate performance measurements in terms of correctness and completion times, we break the high-level task of network exploration down into low-level analytical tasks. 
The selected tasks are based on the well-accepted task taxonomy for graph visualization by Lee et al.~\cite{lee2006task}, who categorize typical low-level graph analysis tasks into four larger categories: 
topology-based tasks, such as finding neighbors of a given node or finding common connections; 
attribute-based tasks, such as finding nodes or edges with certain associated attributes; 
browsing tasks, such as following a given path; and 
overview tasks, where the goal is to estimate values, such as the size of a network, very quickly.



We only investigate tasks that might benefit from local exploration in a detail view, as the overview perspective that serves as the starting point for certain tasks is identical across all interfaces. We therefore do not investigate tasks that are presumably much easier to carry out from an overview perspective, such as estimating the size of a network. 
We selected typical low-level network analytics tasks by Lee et al.~\cite{lee2006task} that fall into the categories of topology-based and browsing tasks. 
Below, we give a detailed description of our experiment tasks in the order in which they were presented to the users. 
Table \ref{tab:tasks} describes how these individual tasks relate to the high-level factors addressed by our hypotheses. 
Figure~\ref{fig:network2D} depicts the individual tasks in the context of one of the networks used in the study, using a 2D force-directed layout for illustration purposes. 





\textbf{Find Neighbor} (FiN): The aim of the task is to probe the ability of users to quickly query the local neighborhood of a hub node in order to find a specific node. Hub nodes for this task had between 14 and 44 neighbors. Users started this task in the detail perspective, from the center of the respective hub node. The task description specified the label of the target node that the user had to find. For reference, the target label was also displayed on the virtual representation of the VR controller. The task was automatically completed when the user \alessio{found and selected the specified} neighbor node with the laser pointer. We measured task performance in terms of completion time. 

\textbf{Find Common Neighbors} (FCN): Inspired by Kwon et al.~\cite{kwon2016study}, the task is to find all common neighbors between a given pair of highlighted nodes. 
Users started this task in the detail perspective, from the center of the first node of the pair. The respective other node of the pair was highlighted and visible when the task was initiated. Each node pair had between one to five common neighbors across all experiment configurations. 
The user had to pick all common neighbors using the laser pointer \johannes{and signaled task completion to the co-present study instructor.} We measured the completion time and logged the user-selected nodes. From the user selection, we then computed the correctness rate, the miss rate, and the false positive rate. 

\begin{table}[]
\centering
\caption {Factors addressed by the hypotheses and their associated task measures. }  
\scalebox{0.95}{

\label{tab:tasks} 
\begin{tabular}{|l|l|l|}
\hline
\textbf{} & \textbf{High-Level Factor} & \textbf{Task Measures}  \\\hline
H1 & Local visual search & FiN completion time \\
&efficiency&FCN completion time\\
&&FCN correctness rate\\
&&END judgement error\\ 
\hline
H2 & Global visual search & FiP completion time\\
&efficiency&FiP path correctness\\
&&FiP path deviation\\ \hline
H3 & Navigation performance & FoP completion time\\ \hline 
H4 & Spatial orientation & SO O$\rightarrow$D angle deviation\\
&&SO D$\rightarrow$D angle deviation\\
&&SO D$\rightarrow$O angle deviation\\ \hline 
H5 & Cyber-sickness & SSQ~\cite{kennedy1993simulator} \\ \hline 
H6 & User preference & Preference ranking \\
&& NASA-TLX~\cite{hart1988development} \\ 
\hline
\end{tabular}
}
\vspace{-15pt}
\end{table}

\textbf{Estimate Node Degree} (END): To assess the users' ability to identify hub nodes, we asked users to estimate the node degree of a specific node, i.e., the number of neighboring nodes. Nodes selected for this task had between 21 and 53 direct neighbors across all experiment conditions. 
Users started in the detail perspective, from the center of the hub node, and had to verbally report their node degree estimate to the experimenter. This task required more navigation in the baseline condition, as the user had to move to a position from where the hub node and its neighborhood could be observed. We therefore did not compare the completion time, and only analyzed the absolute deviation of the user's estimate from the ground truth node degree.

\textbf{Spatial Orientation Overview $\rightarrow$ Detail} (SO O$\rightarrow$D): This task measures spatial orientation after teleportation from the overview to the detail perspective. 
The task was initiated from the overview, where users were presented with two highlighted nodes, i.e., the start and the end node of a path (see Fig.~\ref{fig:overview}). Users had to press a button on the controller to start the task, after which the highlight of the end node was removed and the user was teleported to the detail perspective at the center of the start node. There, users were asked to point in the direction in which they estimated the end node's position. We measured the absolute angular deviation between the ray cast by the user and the ground truth direction vector between the controller position and the end node. 

\textbf{Find Path} (FiP): Finding the shortest path between two given nodes is a common task in immersive network analytics studies \cite{kwon2016study,cordeil2016immersive,huang2017gesture,drogemuller2018evaluating,buschel2019augmented}. This task started from the end position of the previous task (SO O$\rightarrow$D). After indicating the direction estimate, the end node of the path was highlighted again. From the detail perspective at the center of the start node, the user was now required to find the shortest connected path to the end node. The shortest path consisted of either four or five nodes across all experiment conditions. The user had to verbally report the identified nodes along the shortest path between the two highlighted start and end nodes. We counted paths as correct if all reported neighboring nodes were indeed connected by an edge. For correct paths, we computed the path deviation as the number of edges along the user-reported path divided by the number of edges along the shortest path. We also measured the time from task onset until the completed path report. 

\textbf{Follow Path} (FoP): 
In this task, users were asked to follow a highlighted path through the network, as quickly as possible. The task was initiated from the overview perspective, where users were presented with a single highlighted start node. The user had to click on the highlighted node, so that the next node in the task sequence was highlighted. Depending on the interface, the click also initiated a jump to the (clicked) node in the sequence (\egotwo and \egothree), or users had to manually navigate to the proximity of the node (baseline).
We measured the time between clicking the start node and reaching the end node to assess the effect of jumping vs.~flying through the graph. 

\textbf{Spatial Orientation Detail $\rightarrow$ Detail} (SO D$\rightarrow$D): This task measures spatial orientation after network traversal. The task started from the end position of the previous one (FoP). After reaching the end node, users were asked to point the laser pointer on the controller back into the direction of the start node (which was no longer highlighted). We measured the angle deviation as described for SO O$\rightarrow$D. This task is similar to Kwon et al.'s ``recall node locations'' \cite{kwon2016study}. 

\textbf{Spatial Orientation Detail $\rightarrow$ Overview} (SO D$\rightarrow$O): This task measures spatial orientation after teleportation from the detail view to the overview. The task started from the position of the previous task (SO D$\rightarrow$D): After estimating the direction to the start node, users were teleported back to the initial overview position of the SO O$\rightarrow$D task. From this position, users had to indicate the direction of the last node of the path they had followed. We measured the angular deviation of their estimation towards the ground truth direction. 



\subsection{Data}
\label{sec:data}

\begin{figure} 
    \centering
     \includegraphics[width=0.4\textwidth]{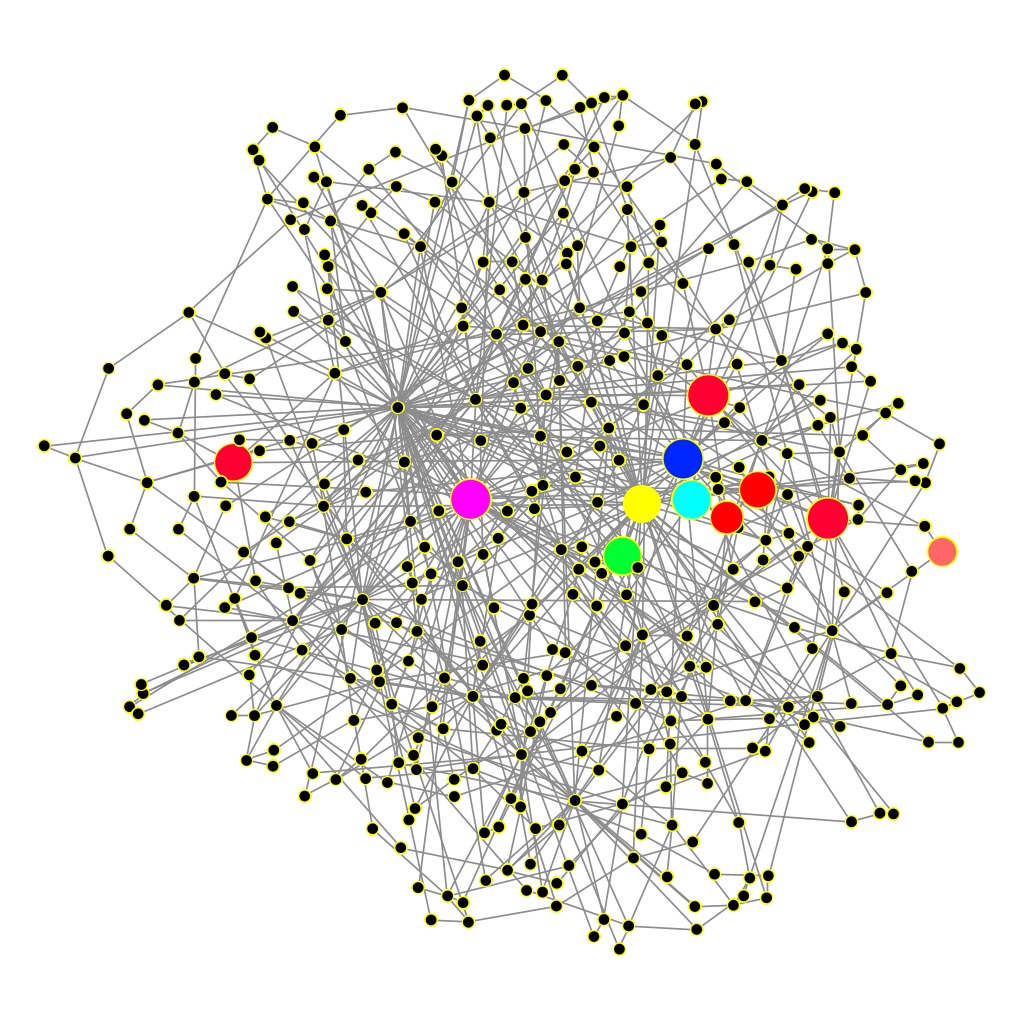}
  \caption{A 2D representation of one of the three networks used for the study, showing color-coded nodes for the different tasks: FiN (yellow = selected, green = target), FCN (purple = selected, green = target), END (blue = selected), FiP (cyan = start, salmon = target), FoP (red nodes).  }
  \label{fig:network2D} 
\end{figure}

In order to achieve meaningful results, we performed our study on networks that resemble real world examples of networks of scientific interest, such as the world-wide-web and authors citation networks~\cite{albert2002statistical}.
Such ``scale-free'' networks, 
have a node degree distribution that follows a power law: the probability $P(k)$ of having a node with degree $k$ decreases exponentially with the law $P(k) \sim k^{-\gamma}$. This results in a small number of high degree nodes (hubs), while the majority of nodes have a low degree.  
When $2<\gamma<3$, the network is an ``ultra-small network'', with a relatively small diameter (i.e., length of the longest shortest path) $d \sim \ln\ln(N)$, and $N$ the size of the vertex set~\cite{cohen2003scale}.
We generated our graphs with the Barab{\'a}si-Albert random graph generator algorithm~\cite{albert2002statistical} (providing $\gamma=3$) using the \textit{NetworkX} Python library~\cite{hagberg2008exploring}. We used six graphs in our study experiments: three with 165 nodes and 326 edges (``small'' in the following), and three with 415 nodes and 826 edges (referred to as ``large'', see Fig.~\ref{fig:network2D}), where the linear density \johannes{(number of edges relative to number of nodes \cite{yoghourdjian2018exploring}) is $2$.}

\subsection{Apparatus}
The presented interfaces and the study framework are implemented as a client-only web application. The code is written in JavaScript using three.js~\cite{threejs} and A-Frame~\cite{aframe}.
The interface between the VR application (i.e., the web-browser) and the VR hardware is handled by SteamVR~\cite{steamvr} that natively supports the HTC Vive.
Our system builds on an open source library for viewing graphs in VR~\cite{forcegraph}. At the time of writing, the core library handles loading, layout, and rendering of the graph, while offering mouse and keyboard navigation.
We extended this core with the presented egocentric interfaces, the study framework (i.e., scripting, scheduling task sequence), as well as to improve the core system's rendering performance.
The most notable extensions concern VR controller support for the HTC Vive, perspective-dependent graph navigation and interaction modes, local layout and rendering adaptions, and support for switching between overview and detail perspectives.

All graph elements had the same color and size, i.e., node diameter and edge thickness.
A single HTC Vive controller was used to handle user inputs via a trigger button, a trackpad and a virtual laser pointer. The laser pointer could be used in all conditions to investigate the direct neighborhood of the closest intersecting node by lowlighting the rest of the graph.
Nodes could be discriminated from each other through unique alphanumeric labels that were always oriented towards the user (see Fig. \ref{fig:teaser}).
In our study, we excluded any sort of artificial, stationary landmarks from the scene to be able to better judge how well users can orient themselves based on the network topology itself (see \textbf{H4}). To create a controlled setting, we 
disabled teleportation between overview and detail view, except for the start (or end) of tasks SO O$\rightarrow$D and D$\rightarrow$O.



\subsection{Study Design}
We used a within-subjects design where \emph{interface} is the independent variable and the individual task measures (see Table~\ref{tab:tasks}) are the dependent variables. 
To counter-balance the order of appearance of the three interface conditions between the participants, as well as the association of graph data sets to the interface conditions, we applied a Graeco-Latin square on these two factors. 
The task order was always presented in the sequence as listed in Section \ref{sec:tasks}. 

\subsection{Procedure}

Before starting the study\footnote{The study was conducted before the global COVID-19 pandemic.}, users had to sign a consent form and fill out a demographic questionnaire.
A general description of the study procedure and of the VR controls were supplied as printouts to participants. 
Each user had to complete the interface conditions in the order specified by the Graeco-Latin square.

Before starting a condition, users were given a short textual description of the condition's interface.
After this, they had to complete a short sequence of tutorial tasks that aimed at familiarizing them with the controls, i.e., navigation and interaction within the current interface. Users could remain in the tutorial until they felt comfortable with the controls and the visual encoding. Until the completion of the tutorial, users were permitted to pose questions about the interface and the study procedure.
For each interface condition, the users then performed the set of tasks described in Section \ref{sec:tasks} twice: once on a small graph, and once on a large graph. Trials using small graphs were treated as training rounds to allow users to familiarize themselves with the tasks and the interface. Only the results of the trials using large graphs were finally analyzed, as discussed in Section~\ref{sec:analysis}, and presented in Section~\ref{sec:results}. 
After each interface condition, users were asked to fill out a Simulator Sickness Questionnaire (SSQ) \cite{kennedy1993simulator}, and a NASA Task Load Index (TLX) \cite{hart1988development} questionnaire. 
At the end of the study, users were asked to rate their preference for the three conditions for solving network analytics tasks. The study was concluded by a semi-structured interview, where users were asked to describe what they liked or disliked about the individual conditions, tasks, and VR network exploration in general \alessio{and report suggestions and improvements.}

\subsection{Participants}

Based on the results of a power analysis after a pilot study, we recruited 25 subjects (19 males, six females) from a local university and an inter-universitary research facility. Participants were aged 23 to 64, with a median age of 31. All participants had normal or corrected to normal vision. 17 participants had a background in computer science, others in physics or finance and economics. Four participants reported never to have played any computer games and not to have any experience with VR. Most users had tried out VR on one or multiple occasions. While 15 users reported to play computer games regularly, six users also play VR games regularly. All 25 participants completed the study. They were compensated with 10 Euros for their time. 

\subsection{Analysis}
\label{sec:analysis}

All obtained responses (completion times, correctness measures, and questionnaire responses) were analyzed individually per task. 
Completion time responses (tasks FiN, FCN, FiP, FoP) were log-transformed to reduce skewness and the effect of outliers~\cite{ratcliff1993methods}, while completion time charts show the untransformed data. All obtained measures were tested for normal distribution. Normally distributed responses were analyzed using a repeated measures~ANCOVA with the three conditions as within-subjects factor and the order of the conditions, given by the Graeco-Latin Square, as covariate. If the normal distribution was violated, we removed outliers (samples higher or lower than $1.5\cdot IQR$). If the filtered responses still did not follow a normal distribution, we performed a Friedman Test. All pairwise post-hoc comparisons were Bonferroni-adjusted. 

\section{Results}
\label{sec:results}

We report the results of the study with respect to the factors listed in Table~\ref{tab:tasks} as $F$- and $\chi^2$-scores, as well as partial $\eta^2$ effect sizes for statistically significant results. The supplemental material contains the results of all performed tests. To put the quantitative results into a qualitative context, we also report user preferences and feedback from a semi-structured interview (see Sec.~\ref{sec:feedback}).  
\vspace{-10pt}
\subsection{Local Visual Search Efficiency}
\label{sec:searchResults}

\begin{figure}[h]
    \centering
    \includegraphics[width=0.8\linewidth]{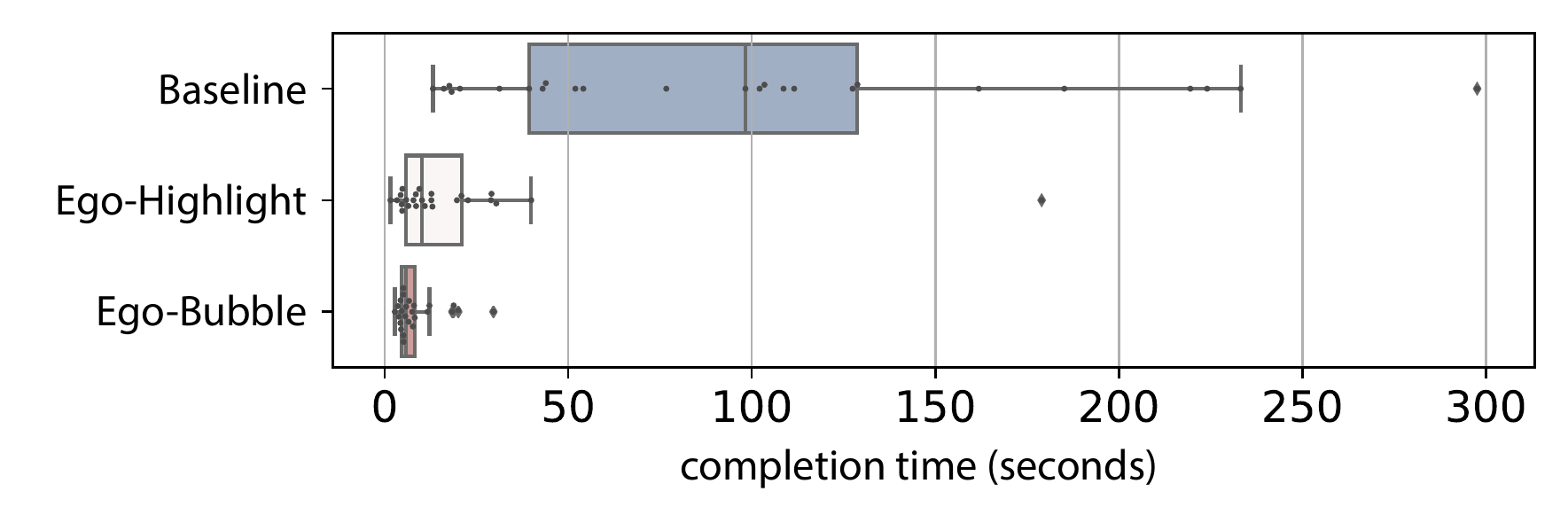}\vspace{-2mm}
    \caption{The completion times in seconds for the FiN task.}
    \label{fig:fin_completion}
\end{figure}

For finding a specified neighbor (FiN), we found that users were significantly slower using the baseline than \egotwo or \egothree ($F_{2,42}=25.722;p<.001;~\eta^2=.551$). On average, users needed around 9 seconds to find the given neighbor using \egothree, 20 seconds using \egotwo, and more than 100 seconds using the baseline (see Fig.~\ref{fig:fin_completion}). 

\begin{figure}[h] 
    \centering
    \subfloat{\includegraphics[width=0.8\linewidth]{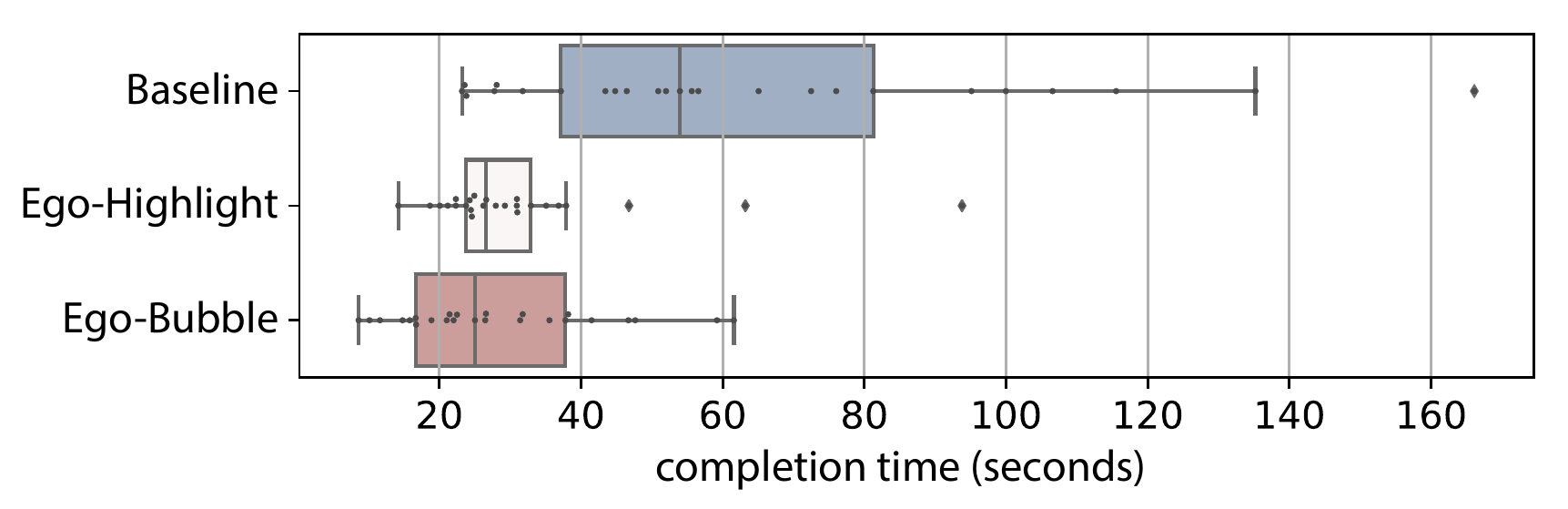}}\vspace{-3mm}
     \subfloat{\includegraphics[width=0.8\linewidth]{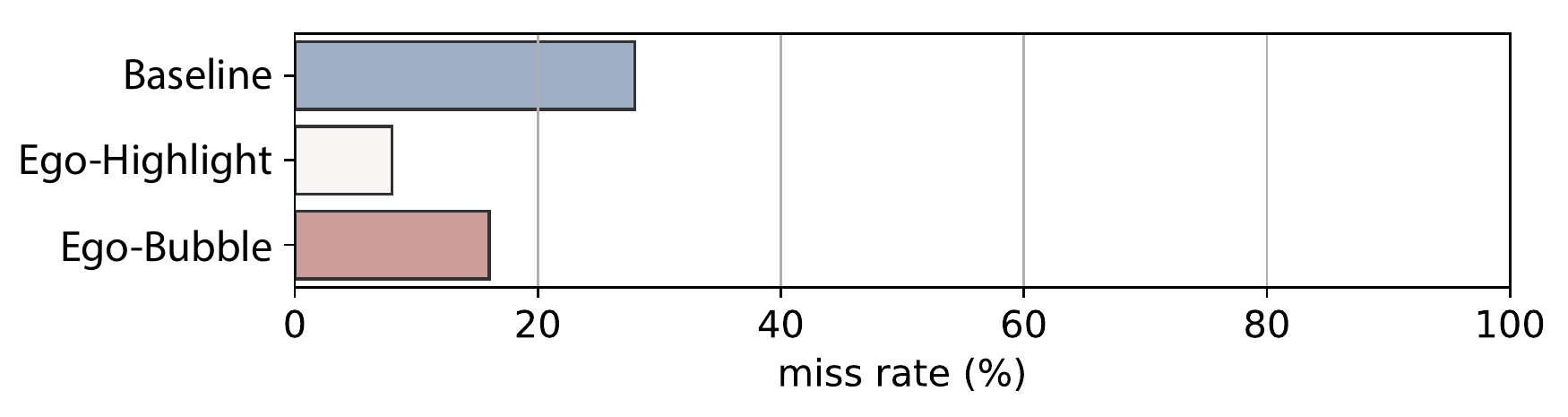}}\\ \vspace{-5mm}
     \subfloat{\includegraphics[width=0.8\linewidth]{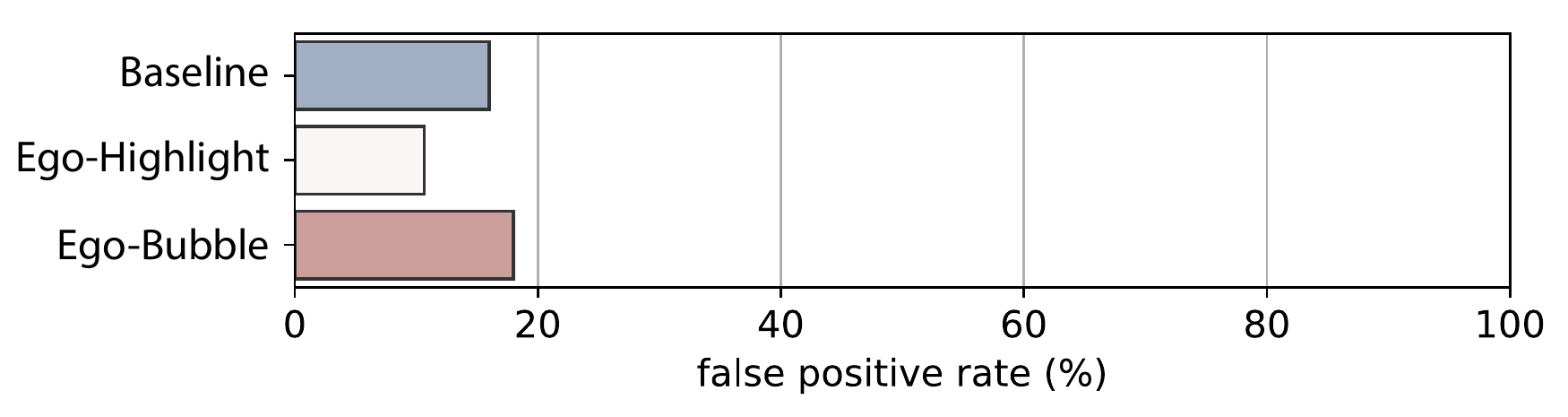}}\vspace{-2mm}
  \caption{Completion times in seconds, miss rates, and false positive rates for the FCN task.}
  \label{fig:fcn_rates} 
\end{figure}

Similarly, we found a large and significant difference in terms of completion time for finding common neighbors (FCN): $F_{1.455,29.101}=20.853;p<.001;~\eta^2=.510$. On average, users required 55 seconds to complete the task using the baseline, which is significantly slower than using \egotwo (32 seconds) \rev{or} \egothree (29 seconds), \rev{which are not significantly different,} as shown in Figure~\ref{fig:fcn_rates}. 
In terms of correctness, we did not find any significant differences for the FCN task. On average, though, users could reach the highest correctness using \egotwo (92\%) and the lowest using the baseline (74\%). The high error rate of the baseline was mainly caused by missed common neighbors (28\% miss rate), while \egothree had the highest false positive rate (18\%), as shown in Figure~\ref{fig:fcn_rates}. 


\begin{figure}[h]
    \centering
    \includegraphics[width=0.8\linewidth]{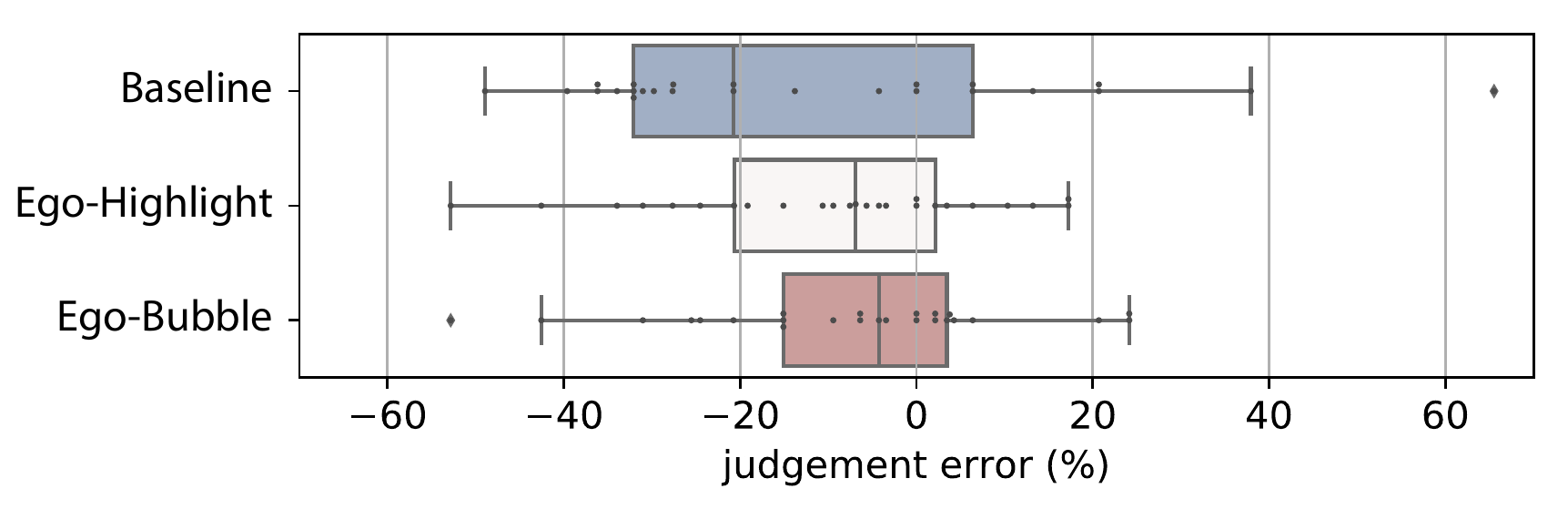}\vspace{-2mm}
    \caption{Judgement error (deviation from ground truth, END task).}
    \label{fig:ed_rates}
\end{figure}

We also found a significant difference between node degree estimation (END) errors ($\chi^2(2)=6.720;p=.035$), where user reports were significantly more deviating from the ground truth node degrees using the baseline (26\%) than when using \egothree (15\%). For all three conditions, users tended to underestimate the number of neighbors rather than overestimate them (see Fig.~\ref{fig:ed_rates}). 

Overall, our hypothesis concerning the effect of egocentrism on local visual search efficiency is partially supported (\textbf{H1}): Indeed, \emph{users are less efficient when analyzing the local neighborhood using the baseline}. The benefit of the \egothree interface in comparison to the \egotwo interface regarding efficiency, however, is rather small and does not reach significance. 

\subsection{Global Visual Search Efficiency}

\begin{figure}[h]
    \centering
    \includegraphics[width=0.8\linewidth]{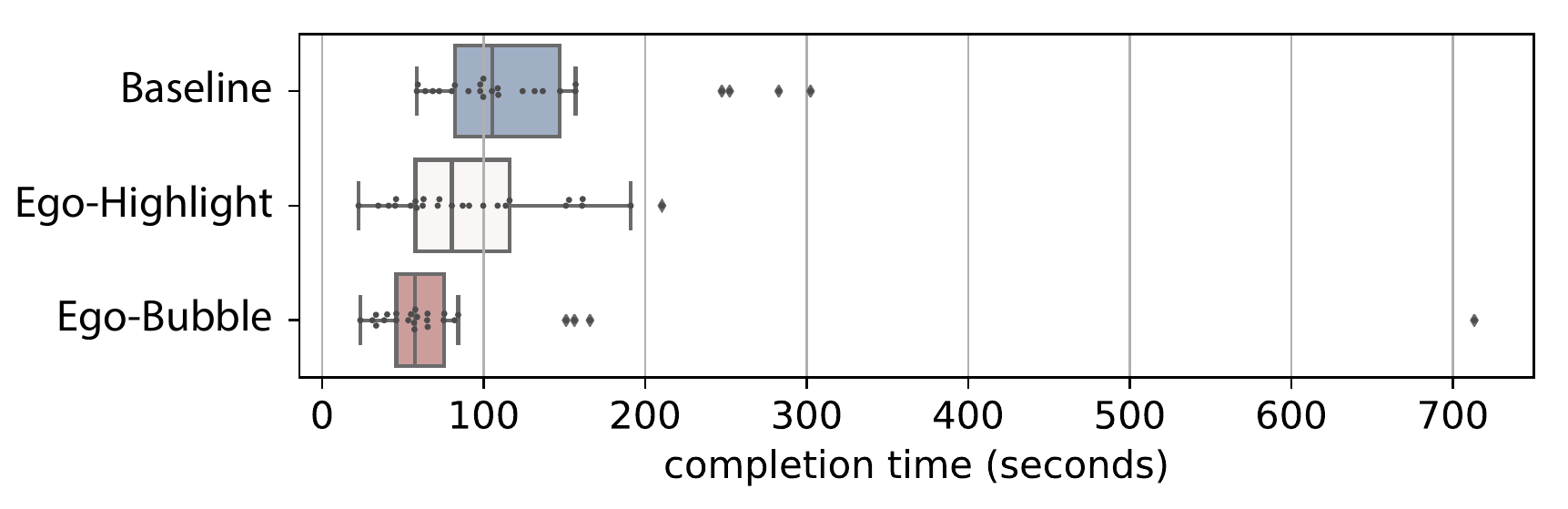}\vspace{-2mm}
    \caption{Completion times in seconds for the FiP task.}
    \label{fig:fip_ct}
\end{figure}

For finding the shortest path, we first removed those users who did not report a correct path in either condition (i.e., they reported non-adjacent nodes or they gave up before reaching the target node). In total, nine responses by six users were incorrect paths, evenly distributed across the three conditions, and nine of the correct paths were longer than the shortest path. The average deviation from the ground truth path length was highest for \egothree (7\%) and lowest for the baseline (3\%), but this difference is not statistically significant. 
There is a large, yet statistically insignificant difference in terms of completion time for those users who reported correct paths: $F_{2,32}=2.956;p=.066;\eta^2=.156$ (see Fig.~\ref{fig:fip_ct}). 

Therefore, our results do not support \textbf{H2}: \emph{Users were only insignificantly more accurate but also slower when trying to find the shortest path between two nodes using the baseline compared to the egocentric interfaces}. 



\subsection{Navigation Performance}
\label{sec:navigationResults}

\begin{figure}[h]
    \centering
    \includegraphics[width=0.8\linewidth]{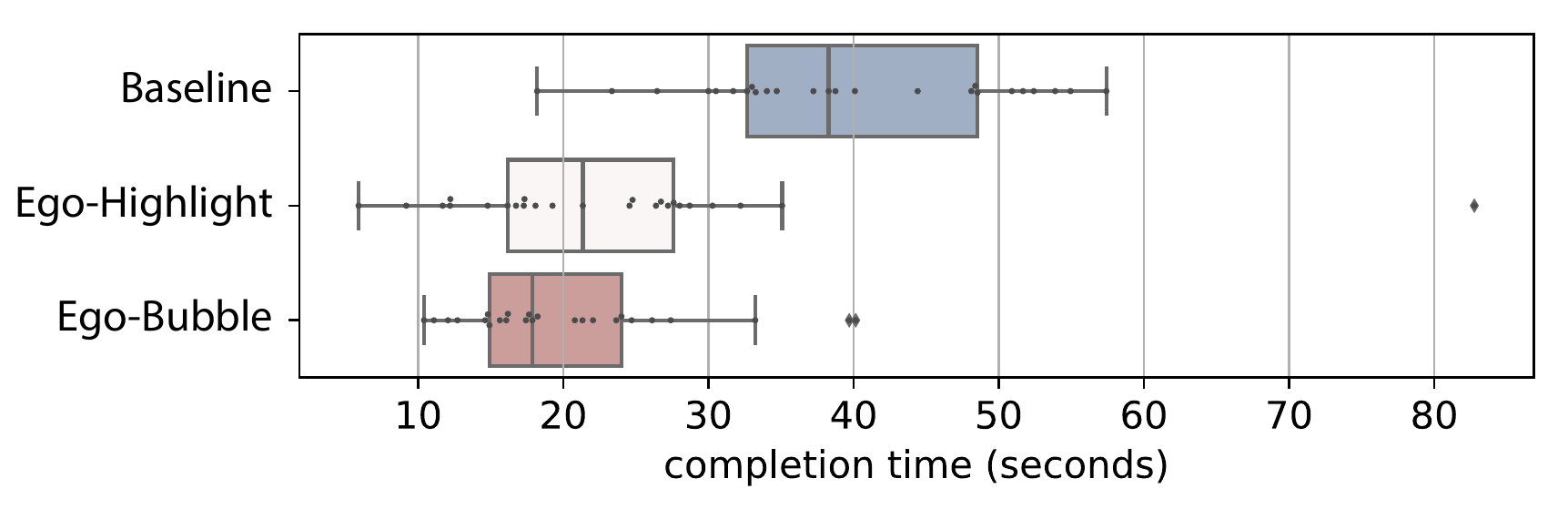}\vspace{-2mm}
    \caption{Completion times in seconds for the FoP task. }
    \label{fig:fop}
\end{figure}

As expected, for the path following task (FoP), we found a large and significant effect of the interface on completion time ($F_{2,44}=49.243;p<.001;p~\eta^2=.691$). \rev{However, the observed speed differences between the interfaces were larger than expected: }
Using the baseline, users required significantly more time to follow the highlighted path (40 seconds, on average) than with both, \egotwo (23 seconds) and \egothree (21 seconds); see Figure~\ref{fig:fop}.

This supports our hypothesis (\textbf{H3}) that \emph{jumping through the network along a highlighted path is faster (almost twice as fast) than flying through it}. \rev{This indicates that egocentrism does not add extra effort for wayfinding.}

\subsection{Orientation}
\label{sec:orientationResults}

 In general, the measured angle deviations for node-direction estimations were quite large, ranging from 19$^{\circ}$ average deviation after jumping from the overview to the detail view (SO O$\rightarrow$D) to an average error of 35$^{\circ}$ when estimating the direction of the start node after following the path in task FoP (SO D$\rightarrow$D). However, for none of the three spatial orientation tasks, we could find any statistically significant difference between the conditions.
 
 
Therefore our hypothesis (\textbf{H4}) that local adaptions introduced through egocentrism lead to decreased spatial orientation in the network is not supported. However, the high angle deviation clearly shows that \emph{explicit spatial orientation cues are required for immersive network exploration, irrespective of the detail interface. }

\subsection{Cyber-sickness}
\label{sec:sicknessResults}

\begin{figure}[ht] 
    \centering
     \subfloat{\includegraphics[width=0.8\linewidth]{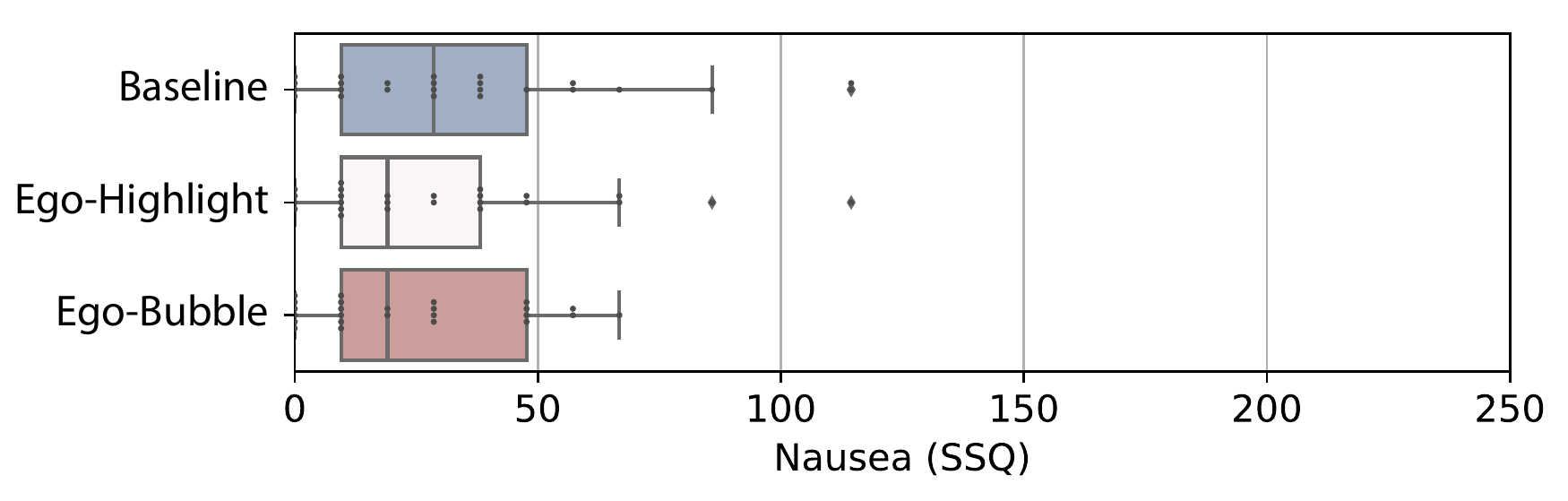}}\\ \vspace{-5mm}
     \subfloat{\includegraphics[width=0.8\linewidth]{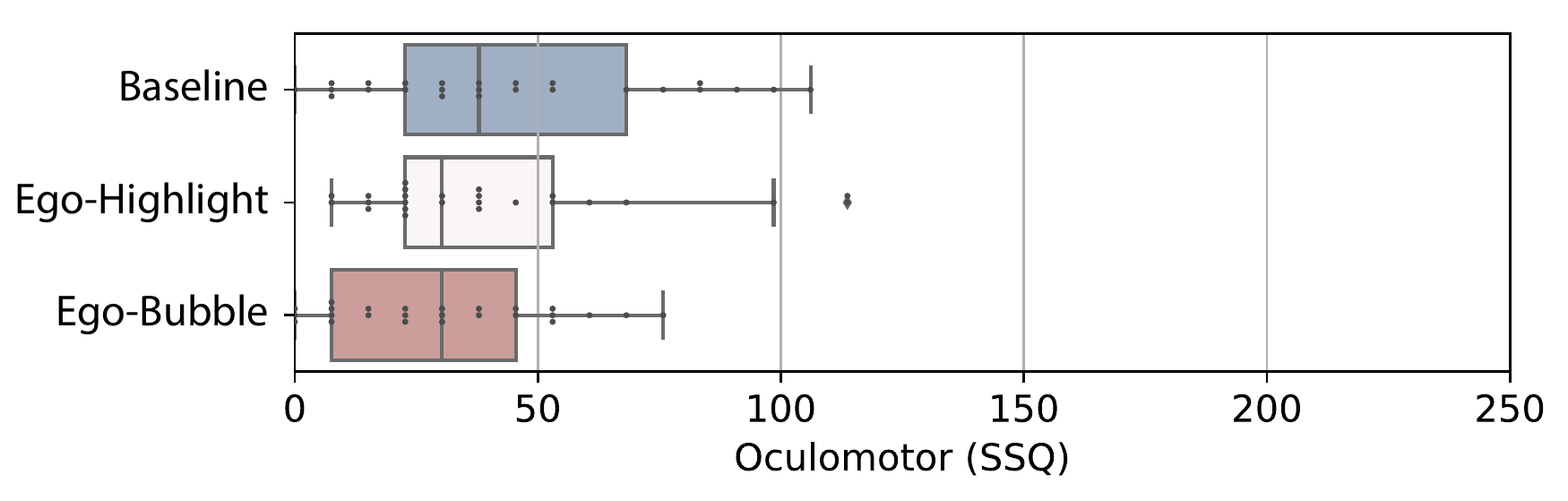}}\vspace{-5mm}
     \subfloat{\includegraphics[width=0.8\linewidth]{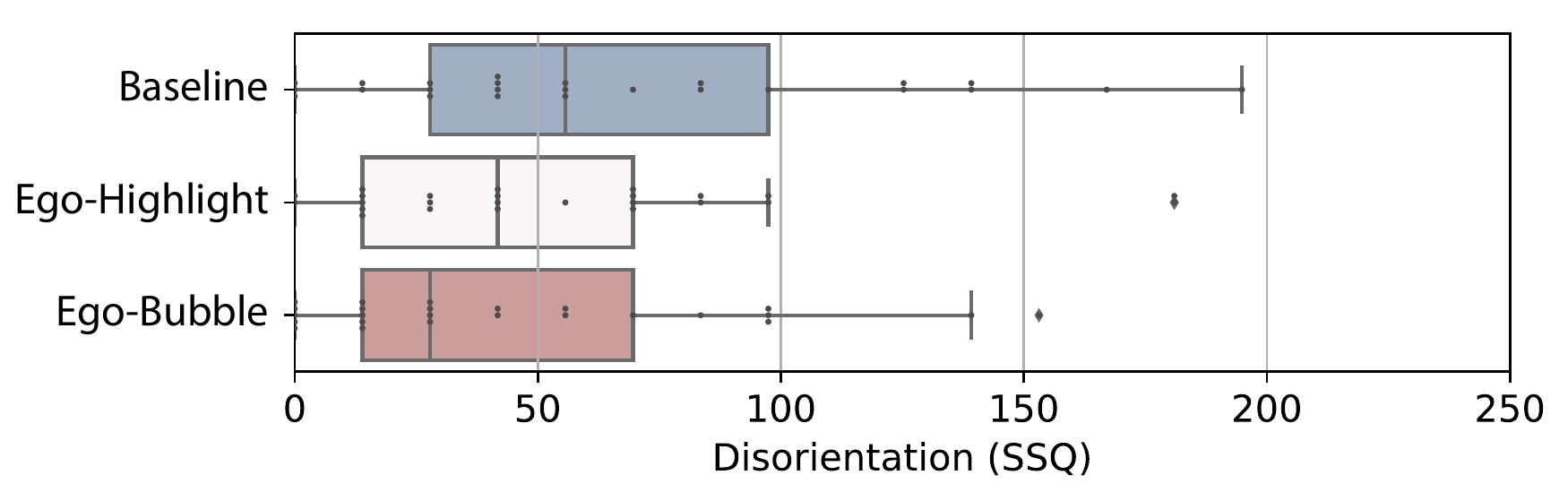}}\vspace{-2mm}
  \caption{SSQ scores related to nausea (top), oculomotor (middle), and disorientation (bottom). }
  \label{fig:SSQ} 
\end{figure}

We derived three metrics from the SSQ, as described by Kennedy et al.~\cite{kennedy1993simulator}: nausea (e.g., sweating and stomach awareness), oculomotor (e.g., fatigue, headache, and eye strain), and disorientation (e.g., vertigo and dizziness). \rev{Note that, due to the within-subjects study design, we compared the scores directly to each other, not to a pre-exposure baseline.} While we did not find any differences between the interface conditions in terms of nausea, we found a significant effect on the oculomotor-related responses ($\chi^2(2)=8.195;p=.017$) and disorientation symptoms ($\chi^2(2)=8.617;p=.013$). Contrary to our hypothesis (\textbf{H5}), post-hoc comparisons for both metrics revealed a significant difference between the baseline and \egothree. On average, \emph{\egothree received slightly lower oculomotor and disorientation symptom scores than the baseline, but the difference between \egotwo and the other two conditions is small and insignificant} (see Fig.~\ref{fig:SSQ}). 

\subsection{User Preferences and Feedback} 
\label{sec:feedback}

We computed an aggregated task load index\rev{, as described by Rubio et al.~\cite{rubio2004evaluation},} to assess the perceived work load of users per condition. The difference between the interface conditions with respect to the perceived task load is significant ($F_{2,48}=6.546;p=.003;~\eta^2=.214$). The baseline received significantly higher task load scores (3.8 on average) than \egothree (3.3) and \egotwo (3.4). 


Analogously to the perceived task load, users also ranked their preference for the baseline considerably lower than for \egotwo and \egothree. Only four users ranked the baseline first; these users appreciated the fact that they were able to move and look around freely and that the interface was very simple. \egotwo and \egothree were ranked first by eleven participants each (with one vote tied between them). Users appreciated \egotwo mainly because of its implicit highlighting of direct neighbors. Three users mentioned that they felt less nausea or vertigo in the \egothree condition, and five described it as ``tidier'' or less cluttered compared to the other conditions. Two users preferring \egotwo stated that they did not find any advantage of the \egothree condition compared to \egotwo. One user preferring the baseline found it hard to orient in the \egothree view, presumably due to the ``nodes moving towards oneself''. 

Users suggested several improvements for the immersive interfaces. Almost half of the users (10) suggested to enable permanent highlighting of user-selected nodes. Three users wanted to have color-coding based on the geodesic distance towards their current location. Three other users suggested to have additional information about the graph nodes and edges (such as their node degree), for instance as a details-on-demand label on the controller. Six users explicitly mentioned that they do not want to be (exclusively) inside the network, but would like to gain some overview from an outside perspective as well. Eight users requested persistent landmarks in the scene to support spatial orientation, and three users would have preferred to have a combination of jumping and flying navigation.
Overall, our hypothesis \textbf{H6} is supported: \emph{Egocentric immersive network exploration is clearly prefered by the users and also leads to a lower subjective task load.}





\section{Discussion}
\label{sec:discussion}


The results of our study support our basic assumption: \alessio{immersive egocentric network exploration interfaces} 
are more efficient and more effective than their non-egocentric counter-part.
Below, we discuss our findings in regard to the aspects investigated in our study. 

\textbf{Visual search in the local neighborhood} is clearly facilitated by a decluttered egocentric interface. For instance, finding a neighbor node with a given label takes around five times as long using the baseline interface. Clearly, the implicit highlighting of neighbors in the egocentric conditions facilitates the identification of connections in the local neighborhood. However, such highlighting is only possible if the user is associated with a chosen node. While the implicit highlighting of neighbors in \egotwo has a large effect, the local layout optimization of the \egothree interface has only a small effect, indicated by slightly less underestimation in the degree estimation task. Interestingly, only a few participants found \egothree ``tidier'', while many users did not notice the difference between the two egocentric interfaces, or they found it irrelevant. 

\textbf{Global visual search for paths} between two distant nodes was considered a hard task. 
The baseline interface was insignificantly more accurate for finding paths, presumably due to edges being always visible and due to unconstrained navigation. 
\rev{From informal user feedback,} we conclude that path finding in a detail perspective is generally not efficient without dedicated interaction support. 

\textbf{Navigation performance} was significantly lower in the baseline than in the egocentric interface conditions, where users could jump between nodes. The navigation performance was effectively doubled when jumping through the network, comparable to results of a prior study using natural scenes~\cite{weissker2018spatial}. Due to the lower navigation speed and manual steering in the baseline, this was expected, yet not to the observed extent. It has to be noted that different flying methods for immersive network exploration have been investigated~\cite{drogemuller2018evaluating,zielasko2016evaluation}, which could slightly improve navigation performance compared to our classic ego-shooter-like navigation in the baseline interface. 

In our study, we observed \textbf{orientation} loss after teleportation between the overview and the detail view for all interfaces. 
Orientation loss after teleportation can be expected~\cite{moghadam2018scene,drogemuller2018evaluating}. However, we also observed considerable loss of orientation after traversing the network in the detail view -- irrespective of whether the user was continuously flying or jumping between nodes. Contrary to our expectations, orientation loss is not higher in the egocentric views than in the baseline condition. \johannes{This is a strong indication that in abstract detail views additional orientation aids, such as static landmarks, are required to support orientation.}

As expected, the baseline received the highest average
\textbf{cyber-sickness} ratings. 
The low cyber-sickness ratings for the \egothree interface were surprising, despite having additional morphing animations during navigation. A \rev{speculative} explanation could be the reduced scene complexity~\cite{so2001metric} and the less pronounced optical flow effect, caused by a constant, farther distance of the neighboring nodes to the camera. \rev{In the future, it might be interesting to investigate the effect of visualization decluttering on the users' well-being during immersive analytics.}
\section{Actionable Insights From Our Study}

Our study allowed us to investigate the effects of egocentrism on immersive network exploration, but it also helped us to identify open challenges that could not be sufficiently resolved by this metaphor alone. Many aspects of network exploration remain challenging in an immersive detail perspective, despite egocentric optimizations: node degrees tend to get underestimated, common neighbors can be missed, non-existing edges are sometimes inferred, and maintaining spatial orientation is difficult.
In an effort to alleviate these issues, we developed an online tool for immersive network exploration, leveraging the following insights from our study~\cite{EgocentricVR}. 
In our web application, overview and detail perspectives can be switched at any point during the exploration.
The overview is represented by a camera object that can be selected with the laser pointer.
The switch is executed through a smooth transition, to avoid potential disorientation after instant teleportation (see Sec.~\ref{sec:discussion}).


\textbf{Permanent node selection:} Almost half of the users suggested such a feature as a means to ``bookmark'' potentially interesting nodes, which could, for instance, help to identify paths. 

\textbf{Geodesic distance encoding:} Path identification was challenging in all interfaces. However, by leveraging information on the node, the user is currently associated with, we can visually encode the geodesic distance to all other nodes to better support such tasks. We encode the distance in a color range from yellow to red: the closer a node's geodesic distance, the closer is its color to red --  clearly showing connected nodes and their path length in respect to the user, which does not necessarily correspond to the Euclidean distance in a force-directed graph layout.  

\textbf{Node Info Display:} As many users requested more quantitative information during their exploration, we added a head-up display attached to the controller displaying information about the current user node's node degree. 

\textbf{Flexible Navigation:} 
In our study, we enabled flying \emph{or} jumping for network navigation. The feedback of some study participants indicates that, in practice, a combination of both modalities could be beneficial for more experienced users. As users are not uniquely associated with a dedicated node during free-flying, our implementation currently applies egocentric adaptions automatically to the closest node in respect to their position within the network. However, informal feedback revealed that fish-eye distortions in combination with free flying induces considerable cyber-sickness. 

\textbf{Landmarks:} Since orientation in the 3D network proved challenging during navigation and after teleportation across all interfaces, we provide a skybox displaying a ground- and sky-plane as a static landmark. 
As additional optional aid we provide visual cues by highlighting all visited nodes when viewing the graph from the overview perspective.
It will be necessary to investigate how such measures will fare in supporting spatial orientation in the future.

\section{Study Limitations and Future Work} 

In our study, we only investigated a small fraction of the vast design space of egocentric network exploration interfaces. Here, we discuss some of the limitations of our study and suggest specific future work. 
We deliberately limited our tasks to those that would potentially benefit from an immersive detail perspective, as the \textbf{overview} perspective was identical in all three tasks. It is clear that many tasks not investigated in this study could be solved more efficiently from an overview perspective, such as estimating the number of communities (similarly as performed by Greffard et al.~\cite{greffard2011visual}) or finding the most central node in the graph. \johannes{Investigating the benefits of egocentric optimizations for network overview- and hybrid-visualizations would be an interesting avenue for future work.
We currently only use the user-selected node (or the user's position in the network) as information on which egocentric optimizations are based on. Further research is needed to evaluate the potential benefits of including other user-centered information, such as the field-of-view or even the user's gaze-focus.}

Our goal was \emph{not} to formally evaluate the benefit of 3D network representations in VR in respect to classic 2D networks. We relied on ample prior evidence suggesting benefits of stereo and immersive 3D networks -- at least as complementary analysis environment -- for path finding, especially with a larger number of nodes~\cite{ware1996evaluating,belcher2003using}, community detection in complex graphs~\cite{greffard2011visual,kraus2019impact}, for distance assessments between 3D points~\cite{wagner2018immersive}, and for a generally improved mental model of the graph~\cite{kotlarek2020study}. 


In our study, we assumed that many users do not have the required space to use a room-scale VR system. Such a setup would allow users to navigate the graph simply by \textbf{walking} through it.
Room-scale navigation potentially reduces the amount of cyber-sickness~\cite{usoh1999walking,langbehn2018evaluation}. 
Yang et al.~\cite{yang2020embodied} recently conducted a study where users could walk in room-sized immersive networks. A formal comparative study between walking, flying, and jumping through a node-link diagram has not yet been conducted. 


According to a survey by Yoghourdjian et al.~\cite{yoghourdjian2018exploring}, the networks we used in our evaluation were very large in terms of the number of nodes but sparse in terms of their \textbf{density}. 
We deliberately limited the number of edges after some informal pilot tests to be able solve local visual search tasks in the baseline condition, which would have been increasingly difficult in denser graphs (cf.~Fig. \ref{fig:teaser}, left). In the future, it will be important to conduct a similar experiment comparing egocentric clutter reducing interfaces in denser networks. Our expectation is that clutter reduction techniques, such as the fish-eye distortion or edge bundling~\cite{holten2006hierarchical, holten2009force}, will become more important with increasing graph density. 


\section{Conclusions}
\johannes{In this work, we promoted \textit{egocentrism} as a metaphor for user-centered optimizations in immersive network exploration interfaces. We introduced two simple instances of egocentrism and systematically investigated their effects in a user study in comparison to a traditional network exploration interface. We obtained initial insights about the metaphor's performance on typical network exploration tasks and potential caveats.
The study showed how interfaces implementing egocentrism considerably increase local visual search efficiency by locally de-cluttering and highlighting the direct node neighborhood. 
While additional local layout optimizations yield comparably fewer advantages, users reported slightly lower symptoms of cyber-sickness compared to the traditional interface. 
Users prefer the egocentric de-cluttered perspective of the network. Sacrificing free navigation is reported to be only a minor trade-off.
We speculate that the benefit of a local distortions will be evident in denser networks, which needs to be studied in the future.


We conclude from our study that egocentric visualization and interaction design that is carefully tailored towards the user's relation to the data can yield significant benefits for immersive detail views in network exploration. 
We therefore see the design and investigation of further effective egocentric metaphors as a great open research challenge in the field of immersive analytics.  
}

\section*{Acknowledgments}
This work is supported by the Austrian Research Agency (FFG) project \#882184. We would like to thank all study participants for their time and effort.

\bibliographystyle{eg-alpha-doi}  
\bibliography{bib}        


\end{document}

%% file: relatedwork.tex
\section{Related Work}

The potential of network analytics in immersive environments has been investigated since the nineties~\cite{crutcher1995managing}, with first experimentation on stereoscopic depth cues in the exploration of networks by Ware and Franck~\cite{ware1996evaluating}. The results of this study suggest that the use of immersive technology could substantially increase the size of the networks a user could grasp. These findings were later confirmed and extended to AR~\cite{belcher2003using}, different VR display technologies~\cite{raja2004exploring,cordeil2016immersive}, and different analytical tasks, such as cluster detection~\cite{greffard2011visual,kraus2019impact} or memorizing previously highlighted nodes~\cite{kotlarek2020study}. 
As VR devices increased in affordability and popularity, so did the number of approaches for network visualization and exploration. For example, Huang et al.~\cite{huang2017gesture} and Erra et al.~\cite{erra2019virtual} presented gesture-based interaction techniques for network exploration and manipulation. In terms of navigation and visualization, immersive network techniques generally rely on an outside overview perspective onto a 3D node-link diagram of the network~\cite{osawa2000immersive,huang2017gesture,yang2020embodied,zielasko2016evaluation,kotlarek2020study,sorger2019immersive}, and/or they allow the user to fly through the node-link diagram using conventional 6DOF controllers~\cite{drogemuller2017vrige,erra2019virtual,sorger2019immersive,yang2020embodied,zielasko2016evaluation}. 

\rev{Immersive displays can also be used for effective node highlighting. For example, Alper et al.~\cite{alper2011stereoscopic} introduce the principle of \textit{stereoscopic highlighting}, where stereoscopic depth is used to draw relevant nodes ``closer'' to the user.} \alessio{Altarawneh et al.~further develop this concept with the \textit{Expand} approach~\cite{altarawneh2014expand} for compound graphs, in which depth encodes the structural relationships between nodes.} 
\rev{The ``deadeye'' technique~\cite{krekhov2019deadeye} renders relevant nodes only on one of the stereoscopic displays, which strongly attracts the user's attention. While stereoscopic highlighting can be more accurate and subtle than static visual cues~\cite{alper2011stereoscopic,krekhov2019deadeye}, these techniques are not applicable for immersive 3D network exploration.} 

Other research efforts target specialized graph presentation and navigation techniques for VR. In their early immersive graph visualization system, Osawa et al.~\cite{osawa2000immersive} introduced ``heat models'' that locally distort the representation of the force-directed graph layout. Nodes in the graph are selected to be the ``foci'', that can absorb or emit heat. Halpin et al.~\cite{halpin2008exploring} displayed graphs in 2D, but allowed users to extrude selected nodes into the third dimension. 
Kwon et al.~\cite{kwon2015spherical} presented spherical graph layouts in VR for seated exploration with mouse interaction, where the entirety of the graph is laid out in the user's field of view. Clutter is reduced by routing edges ``away'' from the user. The authors further demonstrated how spherical projections outperform a 2D layout on typical graph exploration tasks, such as finding common neighbors, path finding, and recall of node locations~\cite{kwon2016study}. Fish-eye techniques are another way to expand the user's field-of-view in immersive environments~\cite{ orlosky2014fisheye,debarba2015embodied}.
A case study by Sorger et al.~\cite{sorger2019immersive} suggests that, for real-world tasks, data scientists appreciate complementary perspectives when analyzing their own, well-known network data: an external overview as stable reference point when starting the network exploration in VR and a detail view as a novel perspective to explore local neighborhoods. 
To evaluate navigation within 3D networks, Drogemuller et al.~\cite{drogemuller2018evaluating} evaluated teleportation and one handed flying against two-handed flying~\cite{mine1997moving} and ``worlds in miniature'' locomotion techniques for local detail view navigation in 3D in networks. 
Their results show that flying is more efficient for finding nodes or paths between nodes in the graph, and that teleportation frequently causes loss of orientation.
In contrast, our proposed approach 
provides data-adaptive egocentric navigation, which 
optimizes the network representation for the user's current perspective. In this work, we present first evidence that this metaphor indeed facilitates local network analysis in room-scale visualizations when compared to traditional interfaces for network exploration in VR.